\documentclass{article}
\usepackage[english]{babel}
\usepackage[utf8]{inputenc}
\usepackage{johd}

\title{Calling for a feminist revolt to decolonise data and algorithms\\ in the age of Datification}

\author{Genoveva Vargas-Solar$^{a}$$^{*}$\\
        \small $^{a}$CNRS, Univ Lyon, INSA Lyon, UCBL, LIRIS, UMR5205, F-69622 Villeurbanne, France  \\
        \small $^{*}$Corresponding author: Genoveva VARGAS-SOLAR; \tt{genoveva.vargas-solar@cnrs.fr}
}

\date{} 

\begin{document}

\maketitle

\begin{abstract} 
Datification, introduced by Kenneth Cukier and Viktor Mayer-Schönberger  (\cite{biltgen2016activity}), refers to the controversial and barbarian possibility of generating digital twins of every being and thing on Earth thanks to sensors, data and algorithms. Digital colonisation and neoliberal processes in a datified world are beyond the limits of the material world, beyond time and geographical location. Big Data and artificial intelligence combined have led to data science representing the scientific and "mechanical" materialisation of datification. 
Scholars like N. Couldry and U. Mejiías  (\cite{couldry2021decolonial}),  P. Ricaurte (\cite{ricaurte2019data}) and S.U. Noble (\cite{noble2018algorithms}) define this data-centred phenomenon as a form of data colonialism  \footnote{{\em Data colonialism combines the predatory extractive practices of historical colonialism with the abstract quantification methods of computing} \cite{couldry2021decolonial}.}. 
Through its problematic epistemic basis, data science steps into human life systematically, massively and without being questioned. 
Global north and south countries see in this project the opportunity to observe, control and tune lives, beliefs, and thoughts without people even being aware. No place is unreachable, and people are progressively formatted to believe and organise their aims towards the ideals of savage capitalism "values" injected by the digital hegemony. There is ongoing colonisation of the minds and expectations with temporalities that see in datafication and the "opportunities" it opens as the ultimate evolution of humanity.

 The impact of datification and algorithms re-creating and reproducing colonial violence promises to be increasingly devastating for people in the margins, women and young children very often.
  Scholars like M. Lampard and S.L. Star (\cite{lampland2009standards}) have problematised the quantification of social phenomena and people's behaviour showing how hegemonic thinking biases the choice of variables representing categories and how they are "engineered" to be processed by machine learning algorithms and numerical methods.
 For example, algorithms adopting supervised learning methods \footnote{(Datasets already labelled with examples of "acceptable" cases are used to train the machine to learn the "good" classification (e.g., Male, Female) } use bi-gender criteria to define classification groups (i.e., male and female, black and white, developed and underdeveloped, rich and poor). 
 Algorithms re-patriarchalise physical and digital spaces operating through hegemony and homogenisation under a dangerous veneer of objectivity and scientific truth"\footnote{Citation in \cite{doi:10.1080/1369118X.2021.1986102} of \cite{noble2018algorithms}, p. 24; \cite{benjamin2019race}, pp. 5–6)}.  They are designed to take the positions of dominant groups. Then their results are adopted as universal positions, marginalising alternatives, erasing differences, and obscuring the particularities encoded in the universal, as studied by P. Dourish and S.D. Mainwaring (\cite{dourish2012ubicomp}). 
 
 The problem is that datification relies on living body territories, creating slavery conditions for maintaining the digital twin. Indeed, the condition in which data are collected and the economic value extracted from "personal" data by companies and governments are dramatic\footnote{N. Couldry and U. Mejías describe this as a "new form of resource appropriation"(\cite{doi:10.1080/1369118X.2021.1986102}).}. The datification cycle is based on 
 data preparation (profiling, cleaning and engineering) that calls for crowdsourcing practices. These new "employment opportunities" enslave humans to validate or annotate (i.e., tag)  data items (e.g. image, video and sound processing). People must work in very precarious forms of continuous production settings if they want to achieve the minimum salary. Poorly annotated items are not paid, the person is sanctioned, the payment per item is reduced, or the access to work offer is restricted for some time. Crowdsourcing is a new form of digital slavery where workers often face solo childcare or long-term unemployment. These repetitive tagging tasks aim to train artificial intelligence algorithms to perform complex classification tasks (identification of nudity in images, images and videos with violent and criminal content). The view of this shocking content generates mental distress. Of course, companies do not provide computers or help with electricity and the implicated costs. 
 
 Still, the practices of datification, for example, crowdsourcing,  numerical classification and ubiquity, are also tools for decolonial actions. These methods can be turned into collective horizontal mechanisms for communities to create decolonial digital spaces that can promote alternative cultural practices, share experiences and knowledge and provide answers against violence, exclusion and generalisation. Feminist perspectives consider gender an essential category to understand the harm
and figure out insurrection actions against datification process that empowers the worst sides of capitalism, and transform it into gore capitalism, the concept introduced by S. Valencia (\cite{valencia2010capitalismo}).

Feminist and women groups, indigenous communities and scholars in the global south/north refusing to adhere to hegemonic datafication program have started to organise and fight back from the inside. The first essential step is to show and problematise technological progress exhibiting the poverty, violence, exclusion, and cultural erase promoted by this "progress". The second step is to promote technology, algorithmic and artificial literacy. Education is critical to learn how to revert and revoke the datified digital twin already colonising all Earth's societies silently and with impunity. It is not the colonisation of body-territories; it goes beyond and occupies humanity's mind's essence, i.e., imagination and imaginary. Against the colonisation of the imaginary, militant groups are imagining and designing alternative algorithms,  datasets collection strategies and appropriation methods. Activist technology groups like Tierra Común \footnote{\url{https://www.tierracomun.net} accessed the 26th July 2022.}, Datactive in the global north\footnote{\url{https://data-activism.net} accessed the 26th July 2022.}, Indigenous Data Sovereignty \footnote{\url{https://www.stateofopendata.od4d.net/chapters/issues/indigenous-data.html} accessed the 26th July 2022.} or La Sandía Digital \footnote{\url{https://lasandiadigital.org.mx} accessed the 26th July 2022.} provide archival servers, or they give people the right to govern their data. These communities flood current platforms and digital spaces with these decolonisation perspectives.
Decolonisation of the meta-verse proposes multi-perspective ways of animating digital and physical bodies' interactions. Activism is willing to create
new bodies, alternative forms of understanding identities, and vital expectations (living in harmony and not occupying and exploiting natural resources). Should decolonisation activisms consider dismantling the meta-verse with its datafication hegemony to imagine less quantitative and more sensitive kaleidoscopic digital territories as multi-verses? Should decolonisation activism instead imagine the extraction of the hegemonic digital ubiquitous meta-verse?




 
\noindent  \end{abstract}



\bibliographystyle{johd}
\bibliography{bib}

\begin{thebibliography}{}

\bibitem [\protect \citeauthoryear {%
Benjamin%
}{%
Benjamin%
}{%
{\protect \APACyear {2019}}%
}]{%
benjamin2019race}
\APACinsertmetastar {%
benjamin2019race}%
\begin{APACrefauthors}%
Benjamin, R.%
\end{APACrefauthors}%
\unskip\
\newblock
\APACrefYearMonthDay{2019}{}{}.
\newblock
{\BBOQ}\APACrefatitle {Race after technology: Abolitionist tools for the new
  jim code} {Race after technology: Abolitionist tools for the new jim
  code}.{\BBCQ}
\newblock
\APACjournalVolNumPages{Social forces}{}{}{}.
\PrintBackRefs{\CurrentBib}

\bibitem [\protect \citeauthoryear {%
Biltgen%
\ \BBA {} Ryan%
}{%
Biltgen%
\ \BBA {} Ryan%
}{%
{\protect \APACyear {2016}}%
}]{%
biltgen2016activity}
\APACinsertmetastar {%
biltgen2016activity}%
\begin{APACrefauthors}%
Biltgen, P.%
\BCBT {}\ \BBA {} Ryan, S.%
\end{APACrefauthors}%
\unskip\
\newblock
\APACrefYear{2016}.
\newblock
\APACrefbtitle {Activity-based intelligence: principles and applications}
  {Activity-based intelligence: principles and applications}.
\newblock
\APACaddressPublisher{}{Artech House}.
\PrintBackRefs{\CurrentBib}

\bibitem [\protect \citeauthoryear {%
Couldry%
\ \BBA {} Mejias%
}{%
Couldry%
\ \BBA {} Mejias%
}{%
{\protect \APACyear {2021}}%
{\protect \APACexlab {{\protect \BCnt {1}}}}}]{%
couldry2021decolonial}
\APACinsertmetastar {%
couldry2021decolonial}%
\begin{APACrefauthors}%
Couldry, N.%
\BCBT {}\ \BBA {} Mejias, U\BPBI A.%
\end{APACrefauthors}%
\unskip\
\newblock
\APACrefYearMonthDay{2021{\protect \BCnt {1}}}{}{}.
\newblock
{\BBOQ}\APACrefatitle {The decolonial turn in data and technology research:
  what is at stake and where is it heading?} {The decolonial turn in data and
  technology research: what is at stake and where is it heading?}{\BBCQ}
\newblock
\APACjournalVolNumPages{Information, Communication \& Society}{}{}{1--17}.
\PrintBackRefs{\CurrentBib}

\bibitem [\protect \citeauthoryear {%
Couldry%
\ \BBA {} Mejias%
}{%
Couldry%
\ \BBA {} Mejias%
}{%
{\protect \APACyear {2021}}%
{\protect \APACexlab {{\protect \BCnt {2}}}}}]{%
doi:10.1080/1369118X.2021.1986102}
\APACinsertmetastar {%
doi:10.1080/1369118X.2021.1986102}%
\begin{APACrefauthors}%
Couldry, N.%
\BCBT {}\ \BBA {} Mejias, U\BPBI A.%
\end{APACrefauthors}%
\unskip\
\newblock
\APACrefYearMonthDay{2021{\protect \BCnt {2}}}{}{}.
\newblock
{\BBOQ}\APACrefatitle {The decolonial turn in data and technology research:
  what is at stake and where is it heading?} {The decolonial turn in data and
  technology research: what is at stake and where is it heading?}{\BBCQ}
\newblock
\APACjournalVolNumPages{Information, Communication \& Society}{0}{0}{1-17}.
\newblock
\begin{APACrefURL} \url{https://doi.org/10.1080/1369118X.2021.1986102}
  \end{APACrefURL}
\newblock
\doi{10.1080/1369118X.2021.1986102}
\PrintBackRefs{\CurrentBib}

\bibitem [\protect \citeauthoryear {%
Dourish%
\ \BBA {} Mainwaring%
}{%
Dourish%
\ \BBA {} Mainwaring%
}{%
{\protect \APACyear {2012}}%
}]{%
dourish2012ubicomp}
\APACinsertmetastar {%
dourish2012ubicomp}%
\begin{APACrefauthors}%
Dourish, P.%
\BCBT {}\ \BBA {} Mainwaring, S\BPBI D.%
\end{APACrefauthors}%
\unskip\
\newblock
\APACrefYearMonthDay{2012}{}{}.
\newblock
{\BBOQ}\APACrefatitle {Ubicomp's colonial impulse} {Ubicomp's colonial
  impulse}.{\BBCQ}
\newblock
\BIn{} \APACrefbtitle {Proceedings of the 2012 ACM conference on ubiquitous
  computing} {Proceedings of the 2012 acm conference on ubiquitous computing}\
  (\BPGS\ 133--142).
\PrintBackRefs{\CurrentBib}

\bibitem [\protect \citeauthoryear {%
Lampland%
\ \BBA {} Star%
}{%
Lampland%
\ \BBA {} Star%
}{%
{\protect \APACyear {2009}}%
}]{%
lampland2009standards}
\APACinsertmetastar {%
lampland2009standards}%
\begin{APACrefauthors}%
Lampland, M.%
\BCBT {}\ \BBA {} Star, S\BPBI L.%
\end{APACrefauthors}%
\unskip\
\newblock
\APACrefYear{2009}.
\newblock
\APACrefbtitle {Standards and their stories: How quantifying, classifying, and
  formalizing practices shape everyday life} {Standards and their stories: How
  quantifying, classifying, and formalizing practices shape everyday life}.
\newblock
\APACaddressPublisher{}{Cornell University Press}.
\PrintBackRefs{\CurrentBib}

\bibitem [\protect \citeauthoryear {%
Noble%
}{%
Noble%
}{%
{\protect \APACyear {2018}}%
}]{%
noble2018algorithms}
\APACinsertmetastar {%
noble2018algorithms}%
\begin{APACrefauthors}%
Noble, S\BPBI U.%
\end{APACrefauthors}%
\unskip\
\newblock
\APACrefYearMonthDay{2018}{}{}.
\newblock
{\BBOQ}\APACrefatitle {Algorithms of oppression} {Algorithms of
  oppression}.{\BBCQ}
\newblock
\BIn{} \APACrefbtitle {Algorithms of Oppression.} {Algorithms of oppression.}
\newblock
\APACaddressPublisher{}{New York University Press}.
\PrintBackRefs{\CurrentBib}

\bibitem [\protect \citeauthoryear {%
Ricaurte%
}{%
Ricaurte%
}{%
{\protect \APACyear {2019}}%
}]{%
ricaurte2019data}
\APACinsertmetastar {%
ricaurte2019data}%
\begin{APACrefauthors}%
Ricaurte, P.%
\end{APACrefauthors}%
\unskip\
\newblock
\APACrefYearMonthDay{2019}{}{}.
\newblock
{\BBOQ}\APACrefatitle {Data epistemologies, the coloniality of power, and
  resistance} {Data epistemologies, the coloniality of power, and
  resistance}.{\BBCQ}
\newblock
\APACjournalVolNumPages{Television \& New Media}{20}{4}{350--365}.
\PrintBackRefs{\CurrentBib}

\bibitem [\protect \citeauthoryear {%
Valencia%
}{%
Valencia%
}{%
{\protect \APACyear {2010}}%
}]{%
valencia2010capitalismo}
\APACinsertmetastar {%
valencia2010capitalismo}%
\begin{APACrefauthors}%
Valencia, S.%
\end{APACrefauthors}%
\unskip\
\newblock
\APACrefYear{2010}.
\newblock
\APACrefbtitle {Capitalismo gore} {Capitalismo gore}\ (\BVOL~158).
\newblock
\APACaddressPublisher{}{Melusina Espa{\~n}a}.
\PrintBackRefs{\CurrentBib}

\end{thebibliography}

\end{document}